# Predicting Cancer Treatments Induced Cardiotoxicity of Breast Cancer Patients

Sicheng Zhou
Institute for Health Informatics
University of Minnesota
Minneapolis, MN, USA
zhou1281@umn.edu

Rui Zhang
Institute for Health Informatics and
College of Pharmacy
University of Minnesota
Minneapolis, MN, USA
zhan1386@umn.edu

Anne Blaes
Department of Medicine
University of Minnesota
Minneapolis, MN, USA
blaes004@umn.edu

Chetan Shenoy
Department of Medicine
University of Minnesota
Minneapolis, MN, USA
blaes004@umn.edu

Gyorgy Simon
Institute for Health Informatics and
University of Minnesota
Minneapolis, MN, USA
zhan1386@umn.edu

*Abstract*—Cardiotoxicity induced by the breast cancer treatments (i.e., chemotherapy, targeted therapy and radiation therapy) is a significant problem for breast cancer patients. The cardiotoxicity risk for breast cancer patients receiving different treatments remains unclear. We developed and evaluated risk predictive models for cardiotoxicity in breast cancer patients using EHR data. The AUC scores to predict the CHF, CAD, CM and MI are 0.846, 0.857, 0.858 and 0.804 respectively. After adjusting for baseline differences in cardiovascular health, patients who received chemotherapy or targeted therapy appeared to have higher risk of cardiotoxicity than patients who received radiation therapy. Due to differences in baseline cardiac health across the different breast cancer treatment groups, caution is recommended in interpreting the cardiotoxic effect of these treatments.

## I. Introduction

Breast cancer is one of the most prevalent and lethal cancers for women. In 2021, an estimated of 281,550 new cases of invasive breast cancer are expected to be diagnosed in women in the U.S. [1]. Cardiotoxicity is a significant problem associated with breast cancer treatments. It is one of the leading causes of death for breast cancer patients, with rates range from 7.4% to 13.3% according to different studies [2,3]. Cardiotoxicity can present acutely or in long term [4-5], and previous studies have shown that breast cancer survivors have a higher cardiac risk [6]. Different cancer treatments (e.g., chemotherapy, radiation therapy and targeted therapy) could cause cardiotoxicity through different mechanisms (Figure 1).

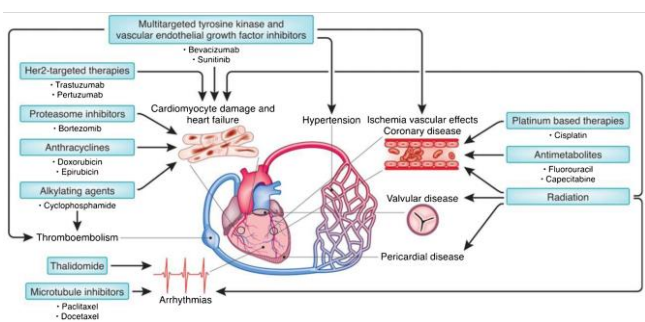

Fig. 1. Different mechanisms for cancer treatment induced cardiotoxicity [2].

The incidence of trastuzumab-related cardiotoxicity could range from 2% to 7% for trastuzumab monotherapy, 2% to 13% for trastuzumab combined with paclitaxel, and can reach to 27% when combined with anthracyclines [7]. Except for causing cardio-related diseases, cardiotoxicity will also delay the treatment of breast cancer which could cause further damage to patients. Thus, accurate prediction of treatment-related cardiotoxicity is paramount for improving patient safety. The lack of comprehensive knowledge about cardiotocixicity is a significant research gap.

Predictive models for various heart diseases have already been developed. A study in 2012 used data from the National Surgical Adjuvant Breast and Bowel Project B-31 to derive a predictive tool for cardiotoxicity. Only two predictors, i.e., age and baseline left ventricular ejection fraction (LVEF) were identified as significant predictors [8].

Ezaz et al. conducted a study that sought to develop a clinical risk score to identify older women with breast cancer who are at higher risk of HF or CM after trastuzumab based on the SEER data [9]. A proportional hazards model was applied to identify candidate predictors of HF and CM and the regression coefficients were then used to construct a risk score. The predictors in the model were age, adjuvant chemotherapy, coronary artery disease, atrial fibrillation or flutter, diabetes mellitus, hypertension, and renal failure. The model was able to classify HF/CM risk into low, medium and high risk group. Baggen, et al. built a multivariate logistic regression model to identify patients with high risk of CHD in clinical settings [10]. 7 variables were included in the final model, i.e., age, congenital diagnosis, NYHA class, cardiac medication, re-intervention, BMI, and NT-proBNP. The developed model was externally validated using C-statistic and obtained a score of 0.78, which indicates good discriminative ability. Currently existing risk models and the identified predictors are not comprehensive enough [11], and only apply to specific types of breast cancer treatments, for instance, patients who take trastuzumab. However, the progression of breast cancer and, accordingly, the treatments could vary a lot among patients. Many potential risk factors, such as characteristics of breast cancer, different treatments and lab results related to cardiovascular factors should be incorporated into the predictive model.

Currently, the evidence-based clinical guidelines for preventing and controlling cardiotoxicity induced by cancer treatments are not specific enough [12], cardiotoxicity risk for breast cancer patients receiving different treatments remains unclear, and there's lack of predictive models of cardiotoxicity developed and evaluated on real-word EHR data. The predicted risk of cardiotoxicity could inform clinicians' cancer treatment choices for patients, and guide the clinicians' to use cardio-protective medications during or after cancer therapy.

The objective of this study is to develop and evaluate comprehensive risk prediction models for cardiotoxicity in breast cancer patients using EHR data. This study is expected to generate practice-based evidence to help better understand the cardiotoxicity for patients with breast cancer receiving various treatments.

## II. METHOD

### A. Data Collection

The dataset for this study contains the breast cancer patients' EHR data extracted from University of Minnesota's Clinical Data Repository (CDR). The ICD-9 and ICD-10 codes were used to identify patients diagnosed with breast cancer between 2011-2020. All patients have treatment records and have minimum follow-up time of 1 year. The patients with prior cancer history except non-melanoma skin cancer or cervical cancer in-situ were excluded for the study. The breast cancer patients are female adults and all patients received one of the radiation therapy, chemotherapy or targeted therapy. Comprehensive factors relevant to heart diseases were collected from the EHR. We chose these factors based on the previous study [13]. We also included some new factors, such as cardiovascular medications, new cancer treatments and lab values. The comprehensive list of predictors under different categories are shown in the Table 1.

TABLE I. COLLECTED VARIABLES FOR BREAST CANCER PATIENTS.

| Category | Variable |
|---|---|
| Outcomes | congestive heart failure (CHF), coronary artery disease (CAD), cardiomyopathy (CM), myocardial infarction (MI) |
| Vitals | systolic blood pressure (SBP), diastolic blood pressure (DBP), body mass index (BMI) |
| Labs | high-density lipoprotein (HDL), low-density lipoprotein (LDL), hemoglobin A1c (Hba1c), troponin, triglyceride, abnormal blood pressure, abnormal blood lipid |
| Pre-conditions | Hyperlipidemia, diabetes, hypertension |
| Cardiovascular related medications | Insulin, Metformin, Statins, ACE inhibitor, Angiotensin II receptor antagonists, Antihypertensive combinations, Vasodilators, Antiarrhythmic, Beta blockers, Calcium blockers |
| Cancer treatments | Radiation therapy, Chemotherapy, Targeted therapy |
| Demographics | Age |

### B. Data Preprocessing

The index date was defined as the first date of any breast cancer treatment (chemotherapy, radiation or targeted therapy), the patients with heart diseases before index date were excluded. Prescriptions of cardiovascular medications and heart diseases (outcomes) were extracted from the follow-up prediod (after the index date). For all other variables, the longitudinal observations before index data were summarized as the value before and closest to the index date. Missing values for continuous variables were imputed either using average values or normal values. For triglyceride, BMI, DBP and SBP, the mean values were used for imputation. For HDL, LDL and Hba1c, the missing values were set to 55 mg/dL, 115 mg/dL and 6% repectively. Several binary variables were also summarized: 1) *antihypertensive_medication*: whether a patient took any antihypertensive medication or not; 2) *antihyperlipedemia_medication*: whether a patient took any antihyperlipedemia medication or not; 3) *abnormal_blood_pressure*: whether a patient had abnormal blood pressure (SBP > 130 mmHg or DBP > 80 mmHg) or not; 4) *abnormal_blood_lipid*: whather a patient had abnormal blood lipid (LDL > 130 mg/dL or HDL < 50 mg/dL or Triglyceride > 150 mg/dL) or not. Each patient received one and only one cancer treatment: 1) Anthracyclines-based chemotherapy: e.g., doxorubicin, epirubicin, daunorubicin, idarubicin, valrubicin; 2) Targeted therapy: limit to Trastuzumab as this is a well-known drug causing cardiotoxicity; 3) Radiation therapy: defined by ICD codes. The *cancer treatment* was transformed into nominal variable with three potential labels (i.e., Radiation therapy, Chemotherapy and Targeted therapy).

### C. Predictive models for heart diseases

The cardiotoxicity cannot be directly measured, we use the risk of four heart diseases (i.e., CAD, CHF, CM and MI) as the measurement of cardiotoxicity. The potential causes of heart diseases could include a series of risk factors such as patients' baseline health levels, pre-conditions and the cancer treatment they received. We applied logistic regression analysis with backwards elimination to construct the predictive models for the four heart diseases separately. The logistic regression model could calculate the probability of an outcome based on the values of risk factors for that event [14]. The predictors include all variables in Table 1 exclude the outcomes. Backwards elimination was applied to select significant predictors after the complete models were built. We applied the 5-fold cross validation to calculate the Area Under the Receiver Operating Characteristics (AUROC) to evaluate the models.

### D. Estimate the treatment effects of cancer treatments on heart diseases

We estimated the average treatment effect (ATE) and the average effect of treatment on the treated (ATT) of breast cancer treatments on the four heart diseases. The ATE and ATT are defined as:

$$ATE = E[Y_i(1) - Y_i(0)]$$

$$ATT = E[Y_i(1) - Y_i(0)|T_i = 1]$$

where $Y_i(0)$ and $Y_i(1)$ are potential outcomes for $T_i = 0$ (a patient receives the cancer treatment) and $T_i = 1$ (a patient not receives the cancer treatment). The ATE and ATT were estimated using logistic models. The outcomes are the risks of four heart diseases. Since every patient received therapy, we set radiation therapy as the reference level and calculate the ATE and ATT of targeted therapy and chemotherapy relative to radiation therapy. We applied 1000 iterations of bootstrap resampling to estimate the confidence interval for ATE and ATT.

*E. Explore patients' baseline health levels for receiving different cancer treatments*

Heart disease events can be a result of cardiotoxicity, but may have also developed independently of breast cancer treatment as a result of patients' pre-existing cardiac risk factors.. We applied logistic regression analysis and the backwards elimination process to explore patients' health level for receiving different treatments. We also set radiation therapy as reference level and built two regression models to compare chemotherapy vs radiation therapy and targeted therapy vs radiation therapy, respectively. The predictors include all baseline labs, vitals and pre-conditions that could reflect the health levels for patients, i.e., *age, sbp, dbp, bmi, ldl, hdl, hba1c, triglyceride, troponin, abnormal_blood_pressure, hypertension, hyperlipidemia, abnormal_blood_lipid and diabetes*. Backward elimination was applied to select significant predictors after the whole models were built. The coefficients of the predictors were summarized to show if patients received different cancer treatments were having different health levels.

*F. Explore patients' medication usage situation for receiving different treatments*

Breast cancer patients may receive medications preventively to protect the cardiovascular system, and to counteract the cardiotoxicity induced by the cancer treatments. Thus, we also explored if patients receiving different breast cancer treatments also received different cardiovascular medications. Similarly, we built two regression models to compare chemotherapy vs radiation therapy and targeted therapy vs radiation therapy, respectively. Tthe outcomes are the three cancer treatments, the predictors include all baseline labs, vitals, pre-conditions, and cardiovascular medications, i.e., *age, sbp, dbp, bmi, ldl, hdl, hba1c, troponin, triglyceride, abnormal_blood_pressure, abnormal_blood_lipid, hyperlipidemia, diabetes, hypertension, metformin, insulin, HMG CoA Reductase Inhibitors, ACE Inhibitors, Angiotensin II Receptor Antagonists, Vasodilators, Antiarrhythmic, Beta_blockers, Calcium_blockers, Diuretics, antihypertensive_medication and antihyperlipidemia_medication*. Backward elimination was applied to select significant predictors after the whole models were built. The coefficients of the predictors were summarized to show if patients received different cancer treatments were taking different medications.

## III. RESULTS

*A. Breast Cancer Patients Cohort*

In total, 3468 breast cancer patients were included in the study. Table 1 listed all the collected variables and the average values or proportions for the variables.

TABLE II. COLLECTED VARIABLES FOR BREAST CANCER PATIENTS.

| Variables | Breast cancer patients (3468) |
|---|---|
| Age (years) | 57.51 |
| SBP (mmHG) | 126 |
| DBP (mmHG) | 74.2 |
| BMI (kg/m2) | 28.7 |
| LDL (mg/dl) | 111.9 |
| HDL (mg/dl) | 65.2 |
| Hba1c (%) | 6.0 |
| Triglyceride (mg/dL) | 128.1 |
| Abnormal_blood pressure (binary) | 43% (1496) |
| Abnormal_lipids (binary) | 17% (592) |
| Diabetes (binary) | 16.52% (573) |
| Hypertension (binary) | 30.25% (1049) |
| Hyperlipidemia (binary) | 23.96% (831) |
| Troponin (binary) | 4.01% (139) |
| Insulin (binary) | 4.30% (147) |
| Metformin (binary) | 4.74% (162) |
| Statins (binary) | 21.69% (742) |
| ACE inhibitor (binary) | 15.52% (531) |
| Angiotensin II receptor antagonists (binary) | 8.97% (307) |
| Vasodilators (binary) | 17.48% (598) |
| Antihypertensive combinations (binary) | 3.63% (124) |
| Antiarrhythmic (binary) | 4.85% (166) |
| Calcium blockers (binary) | 12.34% (422) |
| Antihyperlipidemic (binary) | 1.67% (57) |
| Beta blockers (binary) | 22.27% (762) |
| Chemotherapy (binary) | 21.3% (727) |
| Targeted therapy (binary) | 18.3% (627) |
| Radiation therapy (binary) | 60.4% (2066) |
| CAD (binary) | 4.74% (162) |
| CHF (binary) | 5.64% (193) |
| CM (binary) | 4.82% (165) |
| MI (binary) | 1.40% (48) |

*B. Predictive models for heart diseases*

We constructed four logistics regression models to predict the CAD, CHF, CM and MI separately. The ROC curves and the corresponding AUCs are displayed in Figure 2. The AUC of predictive models for CHF, CAD, CM and MI are 0.846, 0.858, 0.857 and 0.806.

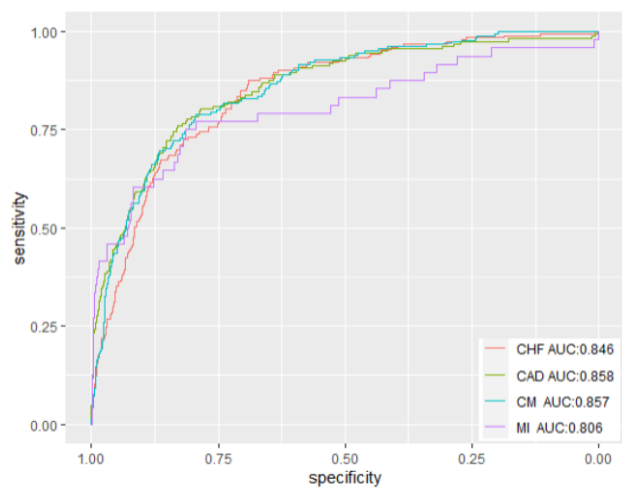

Fig. 2. ROC curves for 4 logistics models to predict the CHF, CAD, CM, MI.

## C. Estimate the treatment effects of cancer treatments on heart diseases

The ATE and ATT of CAD, CHF, CM and MI for chemotherapy and targeted therapy relative to radiation therapy are shown in Table 2. The 95% confidence interval of these scores were calculated using 1000 iterations of bootstrap resampling.

TABLE II.

|  |  | CAD | CHF | CM | MI |
|---|---|---|---|---|---|
| Chemotherapy | ATE | 0.0112± 0.0007 | 0.0545± 0.0007 | 0.0573± 0.0006 | 0.0010± 0.0003 |
|  | ATT | 0.0116± 0.0007 | 0.0559± 0.0007 | 0.0668± 0.0007 | 0.0009± 0.0002 |
| Targeted therapy | ATE | 0.0025± 0.0005 | 0.0574± 0.0007 | 0.0839± 0.0007 | -0.0012 ±0.0002 |
|  | ATT | 0.0029± 0.0007 | 0.0680± 0.0008 | 0.1055± 0.0008 | -0.0017 ±0.0003 |

Compared to radiation therapy (our controls), targeted therapy and chemotherapy both show higher risks for CAD, CHF, CM and MI, however, targeted therapy could result in lower risk for MI.

## D. Explore patients' baseline health levels for receiving different cancer treatments

Variables related to patients' baseline health levels were used to predict the cancer treatments. The coefficients of the significant variables were shown in Table 3.

TABLE III. COEFFICIENTS OF THE SIGNIFICANT VARIABLES TO PREDICT CHEMOTHERAPY, TARGETED THERAPY AND RADIATION THERAPY

| Vars | Coefficients | Standard deviation | Normalized coefficient | p-values |
|---|---|---|---|---|
| Chemotherapy |  |  |  |  |
| age | -0.0593 | 12.2547 | -1.6561 | <0.01 |
| dbp | -0.012 | 10.2974 | -0.2816 | <0.01 |
| bmi | -0.0136 | 6.0845 | -0.1886 | 0.082 |
| ldl | -0.0044 | 19.9402 | -0.1999 | 0.052 |
| hdl | 0.0037 | 28.1748 | 0.2376 | 0.017 |
| hba1c | -0.3288 | 0.4746 | -0.3556 | 0.003 |
| diabetes | 0.2355 | 0.3677 | 0.1973 | 0.086 |
| hypertension | 0.3771 | 0.4608 | 0.3960 | <0.01 |
| triglyceride | -0.0023 | 44.2198 | -0.2318 | 0.048 |
| abnormal blood lipid | 0.6127 | 0.377 | 0.5264 | <0.01 |
| Targeted therapy |  |  |  |  |
| age | -0.0353 | 12.3743 | -1.0336 | <0.01 |
| dbp | -0.0146 | 10.2688 | -0.3548 | <0.01 |
| bmi | -0.0332 | 6.1137 | -0.4803 | <0.01 |
| diabetes | 0.3903 | 0.3694 | 0.3412 | <0.01 |
| triglyceride | -0.0039 | 40.8896 | -0.3774 | <0.01 |
| abnormal blood pressure | -0.3073 | 0.4968 | -0.3613 | <0.01 |
| abnormal blood lipid | 0.4032 | 0.3610 | 0.3444 | <0.01 |

Compared to radiation therapy, the chemotherapy patients tend to have lower *age, dbp, bmi, hba1c, triglyceride* (healthy conditions), but higher rates of *hypertension* and *diabetes* (unhealthy conditions). Targeted therapy patients tend to have lower *age, dbp, bmi, triglyceride, abnormal_blood_pressure* (healthy conditions), but higher rates of *abnormal_blood_lipid* and *diabetes.* (unhealthy conditions).

## E. Explore patients' medication usage situation for receiving different treatments

Variables related to patients' baseline health levels and medications were used to predict the cancer treatments. The coefficients of the significant variables (only medications) were shown in Table 4.

TABLE IV. COEFFICIENTS OF THE SIGNIFICANT MEDICATION VARIABLES TO PREDICT CHEMOTHERAPY, TARGETED THERAPY AND RADIATION THERAPY

| Vars | Coefficients | Standard deviation | Normalized coefficient | p-value |
|---|---|---|---|---|
| Chemotherapy |  |  |  |  |
| age | -0.0586 | 12.2547 | -1.6365 | <0.01 |
| dbp | -0.0105 | 10.2974 | -0.2464 | 0.025 |
| bmi | -0.0156 | 6.0845 | -0.2163 | 0.054 |
| ldl | -0.0046 | 19.9402 | -0.2090 | 0.047 |
| hdl | 0.0068 | 28.1748 | 0.4366 | <0.01 |
| hba1c | -0.3416 | 0.4746 | -0.3695 | 0.003 |
| diabetes | -0.4026 | 0.3677 | -0.3374 | 0.024 |
| Abnormal blood lipid | 0.304 | 0.377 | 0.2612 | 0.015 |
| metformin | 0.7833 | 0.2197 | 0.3922 | 0.002 |
| insulin | 0.432 | 0.2020 | 0.1989 | 0.119 |
| ACE_inhibitor | 0.3174 | 0.3499 | 0.2531 | 0.025 |
| vasodilators | 1.3704 | 0.3567 | 1.1140 | <0.01 |
| Antiarrhythmic | 0.4595 | 0.2060 | 0.2157 | 0.038 |
| Beta blockers | 0.1927 | 0.4057 | 0.1782 | 0.141 |
| calcium blockers | -0.3201 | 0.3215 | -0.2345 | 0.056 |
| Diuretics | 0.4346 | 0.4095 | 0.4056 | <0.01 |
| Targeted therapy |  |  |  |  |
| age | -0.0375 | 12.3743 | -1.0978 | <0.01 |
| dbp | -0.016 | 10.2688 | -0.3887 | <0.01 |
| bmi | -0.0343 | 6.1137 | -0.4961 | <0.01 |
| triglyceride | -0.0032 | 40.8896 | -0.3096 | 0.021 |
| Abnormal blood pressure | 0.3001 | 0.4968 | 0.3527 | 0.016 |
| ACE inhibitor | 0.4462 | 0.3545 | 0.3742 | <0.01 |
| Angiotensin II receptor antagonists | 0.6121 | 0.2849 | 0.4125 | <0.01 |
| vasodilators | 1.1684 | 0.3423 | 0.9462 | <0.01 |
| Antihyperlipidemic | 1.3666 | 0.1281 | 0.4142 | <0.01 |
| Beta blockers | 0.3071 | 0.404 | 0.2935 | 0.023 |
| Antihypertensive medication | -0.2562 | 0.4997 | -0.3029 | 0.022 |

The coefficients indicate that the patients took more cardiovascular medications (e.g., ACE inhibitor, vasodilators, Beta blockers) tend to receive chemotherapy and targeted therapy compared to radiation therapy.

## IV. DISCUSSION

The cardiotoxicity induced by breast cancer treatments is harmful to patients and may delay the cancer treatment. The risks of the cardiotoxicity may vary among different patients. Cardiotoxicity cannot be measured directly, thus we used the four heart diseases as indicators of cardiotoxicity. In this study, we first developed predictive models to estimate the risk of the four heart diseases. However, the risk factors of heart diseases not only include the cancer treatments, but also other factors such as patients' baseline health levels and conditions. We collected a comprehensive set of risk factors to build the predictive models, including patients' vitals, labs, pre-existing conditions and cancer treatments. The AUC were used to evaluated the predictive models. We obtained good AUC for CHF, CAD and CM, which were 0.846, 0.858, 0.857 respectively. The AUC for MI was 0.806, which is relatively low compare to other three models, that's may due to the insufficient cases of MI. There are only 48 (1.4%) MI patients in the cohort, and the highly unbalanced data make limits the performance of the predictive model.

The ATE and ATT of breast cancer treatments on the four heart diseases were estimated using the logistics regression models. The ATE is the treatment's average effect of moving an entire population from untreated to treated status, while the ATT is the average effect of treatment on the patients who actually received the treatment [15]. In our study, there is no patient who did not receive any cancer treatment as the control. Thus, we chose the radiation therapy as the control group (reference level), and computed the ATE and ATT of targeted therapy and chemotherapy relative to radiation therapy. The results indicate that compared to the radiation therapy, the targeted therapy and chemotherapy both could almost result in higher risks for CAD, CHF, CM and MI, except that targeted therapy could result in lower risk for MI (ATE and ATT are negative values). These results indicate that chemotherapy and targeted therapy are more likely to induce heart diseases for breast cancer patients. The radiation therapy has more risk for MI compared to the other three heart diseases, which is consistent with the clinical study [16].

The observed heart diseases could have been caused by cardiotoxicity or could have developed independently due to the patients' pre-existing cardiac risk factors. We further explored whether controls, patients receiving radiation therapy, have different baseline cardiac health levels compared to the patients who received targeted therapy and chemotherapy. Table 3 shows that younger patients, patients with lower *dbp, bmi, ldl, hba1c* and triglyceride and higher *hdl* (which indicates better baseline cardiac health) tend to receive chemotherapy as opposed to radiation therapy; but, at the same time, patients with diagnosis of *diabetes,* and *hypertension* are also more likely to receive chemotherapy. The better lab results despite the higher proportion of diagnoses suggests that these conditions are under control; we cannot exclude the possibility that patients receiving chemotherapy are healthier at baseline. Similarly, patients who had lower *age, bmi, dbp, ldl, triglyceride and abnormal_blood_lipid* (healthier conditions) tend to receive the targeted therapy compared to radiation therapy. Though they also tend to have *diabetes* and *abnormal_blood_pressure* (unhealthier conditions).

Additionally, better outcomes in the radiation therapy (control) group could also be influenced by preventive cardio-protective treatment. Table 4 shows that patients who took *insulin, metformin, ACE inhibitors, antiarrhythmic, vasodilators, beta blockers* and *diuretics* are more likely to receive chemotherapy compared to radiation therapy. Patients *ACE inhibitors, angiotensin II receptor blockers, vasodilators, beta blockers* and *antihyperlipidemic* are likely to receive targeted therapy. Overall, taking the normalized coefficients into account, patients who received chemotherapy and targeted therapy received more cardiovascular medications than the control (radiation therapy) patients.

Some limitations exist in this study. The EHR data only collected from a single healthcare facility, the results need to be further validated using external data. Also, the sample size is limited, especially for the patients with MI. In future we plan to collected more data from multiple healthcare institutes to cross validate and improve the models.

*Conclusion.* In this study we developed predictive models for 4 heart diseases for breast cancer patients based on the cancer treatments along with a comprehensive set of risk factors. Due to the comprehensive set of predictors, the four models achieved high predictive performance, AUCs of are 0.846, 0.857, 0.858 and 0.804, respectively for CAD, CHF, CM and MI, which represents a significant improvement over previous models. Also patients who received chemotherapy and targeted therapy have higher risk of CAD, CHF and CM. Patients in the treatment groups (chemotherapy or targeted therapy) were younger, they had better baseline labs and received more (possibly preventive) cardiovascular medications than patients in the control group (radiation therapy). We adjusted for baseline conditions in our ATT/ATE calculations, but we still recommend caution in drawing conclusions about the expected cardiotoxic effects.

ACKNOWLEDGEMENT

Research reported in this publication was supported by the National Center for Complementary & Integrative Health of the National Institutes of Health under Award Number R01AT009457 (PI: Zhang).